\documentstyle[preprint,aps,version2]{revtex}
\tightenlines

\newcommand{\be}{\begin{equation}}
\newcommand{\ee}{\end{equation}}
\newcommand{\beq}{\begin{eqalignno}}
\newcommand{\eeq}{\end{eqalignno}}
\newcommand{\epem}{\mbox{$e^+e^-$}}

\newcommand{\gev}{{\rm\,GeV}}
\newcommand{\tev}{{\rm\,TeV}}
\newcommand{\fb}{{\rm\,fb}}

\newcommand{\ifb}{{\rm\,fb}^{-1}}

\newcommand{\mz}{M_Z}
\newcommand{\mw}{M_W}
\newcommand{\cosb}{\cos\beta}
\newcommand{\sinb}{\sin\beta}
\newcommand{\tanb}{\tan\beta}
\newcommand{\cosw}{\cos\theta_W}
\newcommand{\sinw}{\sin\theta_W}

\newcommand{\sinsqw}{\sin ^2 \theta_W}
\newcommand{\chc}{\tilde{\chi}^{\pm}}
\newcommand{\chcflip}{\tilde{\chi}^{\mp}}
\newcommand{\chargino}{\tilde{\chi}^{\pm}_1}
\newcommand{\charginotwo}{\tilde{\chi}^{\pm}_2}
\newcommand{\mchargino}{m_{\tilde{\chi}^{\pm}_1}}
\newcommand{\mcharginotwo}{m_{\tilde{\chi}^{\pm}_2}}
\newcommand{\chcp}{\tilde{\chi}^+}
\newcommand{\chcm}{\tilde{\chi}^-}
\newcommand{\chn}{\tilde{\chi}^0}
\newcommand{\LSP}{\chn_1}
\newcommand{\mLSP}{m_{\chn_1}}
\newcommand{\snu}{\tilde{\nu}}
\newcommand{\msnu}{m_{\tilde{\nu}}}

\newcommand{\mslep}{m_{\tilde{\ell}}}

\newcommand{\msq}{m_{\tilde{q}}}

\newcommand{\mpt}{\rlap{$\not$}p_T}

\newcommand{\sigy}{\sigma_Y}

\newcommand{\ds}{\displaystyle}

\newcommand{\mwchi}{M_W^{\chi}}
\newcommand{\mzchi}{M_Z^{\chi}}
\newcommand{\tanbchi}{\tan\beta^{\chi}}
\newcommand{\cosbchi}{\cos\beta^{\chi}}
\newcommand{\sinbchi}{\sin\beta^{\chi}}
\newcommand{\coswchi}{\cos\theta_W^{\chi}}
\newcommand{\sinwchi}{\sin\theta_W^{\chi}}

\newcommand{\gchi}{g^{\chi}}
\newcommand{\gchizero}{g^{\chi^0}}
\newcommand{\dsigr}{d\sigma_R/d\cos\theta}
\newcommand{\dsigl}{d\sigma_L/d\cos\theta}
\newcommand{\onesigma}{1\sigma}
\newcommand{\npts}{N_{\rm pts}}
\newcommand{\achi}{A^{\chi}}
\newcommand{\amctot}{\Delta A_{\rm MC}^{\rm tot}}
\newcommand{\amcstat}{\Delta A_{\rm MC}^{\rm stat}}
\newcommand{\asys}{\Delta A^{\rm sys}}
\newcommand{\emctot}{\Delta \eta_{\rm MC}^{\rm tot}}
\newcommand{\emcstat}{\Delta \eta_{\rm MC}^{\rm stat}}
\newcommand{\esys}{\Delta \eta^{\rm sys}}
\newcommand{\afbr}{A_R^{\chi}}
\newcommand{\afbl}{A_L^{\chi}}

\newcommand{\afbhad}{A^{\rm had}}

\newcommand{\half}{\frac{1}{2}}
\newcommand{\third}{\frac{1}{3}}
\newcommand{\twothird}{\frac{2}{3}}

\newcommand{\rarr}{\rightarrow}

\begin{document}
\draft

\pagestyle{empty}
\preprint{SLAC--PUB--6654}
\medskip
\preprint{LBL--36101}
\medskip
\preprint{UH--511--804--94}
\medskip
\preprint{hep-ph/9502260}
\medskip
\preprint{February 1995}
\medskip
\preprint{T/E}

\begin{title}
Testing Supersymmetry at the Next Linear Collider
\end{title}

\author{
J. L. Feng
\thanks{Work supported by the Department of Energy, contract
DE--AC03--76SF00515.}
\thanks{Work supported in part by an NSF Graduate Research Fellowship.}
\addtocounter{footnote}{-2}
and M. E. Peskin \footnotemark
\addtocounter{footnote}{1}
}

\begin{instit}
Stanford Linear Accelerator Center\\
Stanford University, Stanford, California 94309
\end{instit}

\vspace*{-0.15in}

\author{
H. Murayama
\thanks{Work supported by the Director, Office of Energy Research,
Office of High Energy and Nuclear Physics, Division of High Energy
Physics of the U.S. Department of Energy under contract
\mbox{DE--AC03--76SF00098}.} }

\begin{instit}
Theoretical Physics Group, Lawrence Berkeley Laboratory \\
University of California, Berkeley, California 94720
\end{instit}

\vspace*{-0.15in}

\author{
X. Tata
\thanks{Work supported by the Department of Energy, contract
DE--FG--03--94ER40833.} }

\begin{instit}
Department of Physics and Astronomy \\
University of Hawaii, Honolulu, Hawaii 96822
\end{instit}


\begin{abstract}

\vspace*{-0.5in}

Up to now, almost all discussion of supersymmetry at future colliders
has been concerned with particle searches.  However, if candidates for
supersymmetric particles are found, there is much more that we will
want to know about them. Supersymmetry predicts quantitative relations
among the couplings and masses of supersymmetric particles. We discuss
the prospects for testing such relations at a future $e^+e^-$ linear
collider, using measurements that exploit the availability of polarized
beams.  Precision tests from chargino production are investigated in
two representative cases, and sfermion and neutralino processes are
also discussed.

\vspace*{0.15in}

\centerline{(Submitted to Physical Review {\bf D})}

\end{abstract}


\setcounter{footnote}{0}
\newpage
\pagestyle{plain}
\narrowtext

\section{Introduction}
\label{sec:Introduction}

The phenomenological predictions of supersymmetry (SUSY) may be divided
into three categories: (I) reflections of the supersymmetric Lagrangian
in standard model phenomenology, including relations among the gauge
coupling constants from SUSY grand unification and the presence of a
heavy top quark and a light Higgs scalar; (II) the prediction of new
particles with the correct spin and quantum number assignments to be
superpartners of the standard model particles; and (III) well-defined
quantitative relations among the couplings and masses of these new
particles. While the predictions of (I) are of great interest, their
verification is clearly no substitute for direct evidence.  The
discovery of a large number of particles in category (II) would be
strong support for SUSY. On the other hand, the most compelling
confirmation of SUSY would likely be the precise verification of the
relations of category (III). This would be especially true if,
initially, only a small set of candidate SUSY partners are observed.

Most discussions of supersymmetry at  future high-energy colliders have
concentrated single-mindedly on the question of particle searches. From
one point of view, this is reasonable, because the existence of SUSY
partners is unproven and this a prerequisite for any  further analysis.
On the other hand, the discovery of the first evidence for SUSY --- or
for any other theoretical extension of the standard model --- will
begin a program of detailed experimental investigation of the new
sector of particles required by this extension.  This investigation
will need to be carried out with the same experimental tools that were
used to make the original discovery.  Thus, it is not only reasonable
but also crucial, as we plan for the colliders of the next decade, to
ask how any new physics that might be discovered can be examined in
detail at these machines.

Supersymmetry provides a particularly interesting subject for studies
of the detailed analysis of physics beyond the standard model. SUSY
models are weakly coupled, so their consequences can be worked out
straightforwardly using perturbative computations.  At the same time,
SUSY models depend on a large number of unknown parameters, and
different choices for these parameters yield qualitatively different
realizations of possible new physics.  Thus, the phenomenology of SUSY
is quite complex. Eventually, if SUSY does give a correct model of
Nature, the colliders of the next generation will be expected to
determine the SUSY parameters, and their values will become clues that
take us a step closer to a fundamental theory.  We suggest that similar
complexity should be found in any realistic extension of the standard
model, and that similar investigations will be needed to understand the
next, more fundamental, level.

One consequence of the complexity of the parameter space of SUSY models
is that it is not trivial to identify experimentally the specific
quantities which are related by supersymmetry. Faraggi, Hagelin,
Kelley, and Nanopoulos \cite{FHKN}, Martin and Ramond \cite{Martin},
and Kawamura, Murayama, and Yamaguchi \cite{KMY} have discussed in
general terms the exploration of the spectroscopy of supersymmetry
partners, and the latter two groups have suggested particular mass
relations which test supersymmetry independently of more detailed
hypotheses.  These tests are very ambitious, since they require mass
measurements for the heaviest and most elusive particles of the
superspectrum --- the squarks, the heaviest partners of the Higgs and
gauge bosons, and the sneutrino --- at the 1\% level.  In these papers,
very little attention was given to the question of how these
experiments will be done. In this paper, we will present some
alternative tests of supersymmetry that involve only the lightest
observable states of the superspectrum, and we will argue that these
should be straightforward to carry out at colliders of the next
generation.

Our tests will exploit the advantages of the proposed Next Linear
Collider (NLC), a  linear $\epem$ collider with $\sqrt{s} = 500 \gev$
and a design luminosity of $50\ifb$/year \cite{NLCDesign}.  This
machine has already been shown to be a powerful tool for probing new
physics \cite{Finland,JLC,DESY,Hawaii}. In particular, previous work
has shown that such a machine provides an excellent environment for
measuring SUSY parameters under the assumption that newly discovered
particles are sparticles
\cite{LeNLC,Fujii,Orito,Vandervelde,Kon,squark,INStalk}.   In this
paper, we add to this body of work by showing how to test this
assumption.  Our analysis will take into account the relation of
observable properties of the final state to the underlying reaction; as
in the earlier NLC studies, we will be helped dramatically by the clean
experimental environment expected at this machine. In addition, the
expected availability of highly polarized electron beams should provide
a powerful diagnostic tool.

This study will be conducted in the context of the minimal
supersymmetric standard model (MSSM). It is a reasonable expectation
that charginos --- the mixed superpartners of $W$ bosons and charged
Higgs bosons --- will be among the lightest supersymmetric states, and
that these will be accessible to the NLC. Thus, we concentrate here on
tests of supersymmetry that involve the properties of charginos. The
crucial problem we will face is that the mass eigenstates of charginos
are in general a mixture of weak eigenstates, and their mixing pattern
must be resolved before the quantitative implications of supersymmmetry
become clear.  To understand the experimental aspects of chargino
reactions needed in this study, we have studied simulations of chargino
production and decay using the parton-level Monte Carlo event generator
of Feng and Strassler \cite{LEPIIstudy}.

The outline of this paper is as follows: In Sec.~\ref{sec:MSSM} we
review the properties of charginos within the MSSM and state our
assumptions.  In Sec.~\ref{sec:Charginos} we divide the parameter space
into characteristic regions.  In Secs.~\ref{sec:Mixed} and
\ref{sec:Gaugino}, we present two different strategies for
supersymmetry tests in two of these regions and analyze the
experimental prospects for these tests in particular cases studes. In
Sec.~\ref{sec:Others}, we comment on other possible supersymmetry tests
involving the properties of matter scalars and neutralinos. We present
our conclusions in Sec.~\ref{sec:Conclusions}.

\section{The MSSM and our Assumptions}
\label{sec:MSSM}

Though our goal in the studies reported here is to test supersymmetry,
we cannot begin without narrowing the phenomenological context. SUSY
can, in principle, be realized in many ways.  Here we assume that the
observed particle content and qualitative phenomenology is that of the
minimal supersymmetric extension of the standard model (MSSM), with
conserved R-parity and therefore a stable lightest supersymmetric
particle (LSP).  This is the set of assumptions that is associated with
the most commonly studied missing energy signatures for the discovery
of candidate supersymmetric particles.  R-parity conservation and the
existence of only two Higgs doublets will be our two primary
assumptions, and will be essential for much of the following analysis.
We will also incorporate some minor additional restrictions for
simplicity.  In this section, we detail these assumptions and define
the basic set of parameters.  A more detailed presentation of the MSSM
can be found in many reviews \cite{Reviews}.

The MSSM includes matter superfields and two Higgs doublet superfields
$\hat{H}_1$ and $\hat{H}_2$, which give masses to the isospin
$-\frac{1}{2}$ and $\frac{1}{2}$ particles, respectively.  These two
superfields are coupled in the superpotential through the term $ - \mu
\epsilon_{ij} \hat{H}_1^i \hat{H}_2^j$, and the ratio of the two Higgs
scalar vacuum expectation values is defined to be $\tanb \equiv \langle
H^0_2 \rangle / \langle H^0_1 \rangle$. The MSSM also contains soft
SUSY breaking terms \cite{Savas,Girardello}, which are parametrized by
masses $m_i$ for the scalar multiplets and masses $M_1$, $M_2$, and
$M_3$ for the U(1), SU(2), and SU(3) gauginos.  In addition, there are
cubic couplings (``$A$ terms'') of Higgs scalars and sfermions. With
the assumptions that we will make below, our study will be insensitive
to the parameters entering through the $A$ terms.

The Higgsinos and electroweak gauginos of the MSSM mix to form two
charginos and four neutralinos. In two-component spinor notation, the
chargino mass eigenstates are $\tilde{\chi}^+_i = {\bf V}_{ij}\psi^+_j$
and $\tilde{\chi}^-_i = {\bf U}_{ij}\psi^-_j$, where $(\psi^{\pm})^T =
(-i\tilde{W}^{\pm}, \tilde{H}^{\pm})$ and, by convention, $\mchargino <
\mcharginotwo$. The matrices $\bf V$ and $\bf U$ diagonalize the mass
terms

\be
(\psi ^-)^T {\bf M}_{\chc} \psi^+ + {\rm h.c.} \ ,
\ee
where

\be\label{chamass}
{\bf M}_{\chc} = \left( \begin{array}{cc}
 M_2                    &\sqrt{2} \, \mw\sinb  \\
\sqrt{2} \, \mw\cosb   &\mu                    \end{array} \right) \ .
\ee
Ignoring some subtleties in this diagonalization having to do with
negative mass values and the ordering of the eigenstates (see, for
example, the first reference in \cite{Reviews}), $\bf V$ and $\bf U$
are orthogonal  matrices which can be parametrized by rotation angles
$\phi_+$ and $\phi_-$.  For $\phi_{\pm} = 0$, the chargino $\chargino$
is pure gaugino, and for $\phi_{\pm} = \frac{\pi}{2}$, $\chargino$ is
pure Higgsino. The neutralino mass eigenstates are $\chn_i = {\bf
N}_{ij}\psi^0_j$, where $(\psi^0)^T = (-i\tilde{B},-i\tilde{W}^3,
\tilde{H}^0_1, \tilde{H}^0_2)$, and ${\bf N}$ diagonalizes the mass
terms

\be
\frac{1}{2} (\psi ^0)^T {\bf M}_{\chn} \psi^0 + {\rm h.c.} \ ,
\ee
where

\be\label{neumass}
{\bf M}_{\chn} =
 \left( \begin{array}{cccc}
M_1             &0              &-\mz\cosb\sinw &\mz\sinb\sinw  \\
0               &M_2            &\mz\cosb\cosw  &-\mz\sinb\cosw \\
-\mz\cosb\sinw  &\mz\cosb\cosw  &0              &-\mu           \\
\mz\sinb\sinw   &-\mz\sinb\cosw &-\mu           &0     \end{array}
\right) \ .
\ee

To reduce the large number of arbitrary parameters, we follow
Ref.~\cite{LEPIIstudy} in introducing some additional assumptions.
These assumptions are primarily phenomenologically motivated, and,
where possible, we avoid assumptions based solely on grand unified
theories (GUTs) and supergravity theories. As noted above, we assume
R-parity conservation and the presence of a stable LSP, which we
identify as the lightest neutralino $\LSP$. In addition, we will ignore
the intergenerational mixing in the quark and sfermion sectors, and we
will assume that $CP$-violating phases in the SUSY parameters are
negligible. We will also assume that one-loop effects do not introduce
large and qualitatively new dependences on SUSY parameters. If these
effects are large but may be absorbed by redefinitions of the tree
level parameters, our analysis can be applied with only minor
modifications. The assumptions listed above will be in effect
throughout this study. Additional conditions that are appropriate to
the study of specific processes and scenarios will be given below.

\section{The Parameter Space of Charginos}
\label{sec:Charginos}

In many supersymmetric models, charginos are the lightest observable
sparticles, and we now consider the possibilities for tests of SUSY
{}from chargino production.  As we are interested in what may be
learned from the chargino signal, we will make, in this and the
following two sections, the additional assumptions that gluinos,
sfermions, and the Higgs scalars $H^0$, $A^0$, and $H^{\pm}$ are beyond
the kinematic reach of the NLC.  Neutralino masses must be comparable
to chargino masses, and below we will address the problem of removing
neutralino backgrounds to the chargino signal. If a number of
additional SUSY signals are available at NLC energies, their detection
would be exciting in their own right, and would make possible the
measurement of several sparticle masses.  However, the procedure we
outline below for measuring chargino couplings would not directly
apply.  Since we think it would be somewhat optimistic to expect a
plethora of sparticles to be accessible at NLC energies, we have not
explored this scenario further.

The analysis of chargino pair production and decay is discussed in
detail in Ref.~\cite{LEPIIstudy}; here we will only summarize the most
important qualitative features of this process. Using the picture of
chargino production derived from this analysis, we will divide the
parameter space into characteristic regions.  In the following two
sections, we will define and analyze tests of supersymmetry which rely
on the particular characteristics of the chargino in each of these
regions.

Though the observables we will discuss involve only the chargino pair
production cross section, the problems of experimental detection of the
chargino signal necessarily bring in parameters of the chargino decay
processes.  We simplify our treatment of these processes in the
following way: motivated by $\mu \rarr e \gamma$ and flavor changing
neutral current constraints \cite{FCNC}, we assume that all left-handed
sleptons of different generations are roughly degenerate (to within,
say, 20 GeV) with mass $\mslep$, and the left-handed squarks of the
first two generations are roughly degenerate with mass $\msq$. In fact,
chargino events are usually insensitive to all other sfermion masses.
Decays through third generation squarks are suppressed because, for NLC
energies, the mass difference $\mchargino - \mLSP$ is almost always
less than the top quark mass. For the remaining sfermions, the
right-handed sfermion diagrams are suppressed by Higgs couplings
$m_f/\mw$ and are negligible.

With these assumptions, there are only six parameters that enter the
complete description of chargino pair-production: $\mu$, $M_2$,
$\tanb$, $M_1$, $\mslep$, and $\msq$. We do not assume gaugino mass
unification, and so $M_2$ and $M_1$ are unrelated. With an $e^-_L$
beam, $\chcp_1 \chcm_1$ production occurs through the $s$-channel $Z$
and $\gamma$ diagrams and the $t$-channel $\snu_e$ exchange diagram of
Fig.~\ref{fig:productiondiagrams}, and so the left-handed differential
cross section is governed by four parameters:

\be
\frac{d\sigma_L}{d\cos\theta} \left( e^-_L e^+ \rarr \chcp_1 \chcm_1
\right) = \frac{d\sigma_L}{d\cos\theta} \left( \mu, M_2, \tanb, \mslep
\right) \ .
\ee
In the case of an $e^-_R$ beam, the $\snu_e$ diagram is absent, and so
the right-handed differential cross section is dependent on only the
first three parameters:

\be
\frac{d\sigma_R}{d\cos\theta} \left( e^-_R e^+ \rarr \chcp_1 \chcm_1
\right) = \frac{d\sigma_R}{d\cos\theta} \left( \mu, M_2, \tanb \right)
\ .
\ee
Charginos decay to the LSP either leptonically through $W$ bosons or
virtual sleptons,

\be
\chcp_1\rightarrow (\LSP {W^+}^{(*)}, {\tilde{\ell}}^* \nu,
\bar{\ell} {\tilde{\nu}}^*)\rightarrow \LSP \bar{\ell} \nu \ ,
\ee
or hadronically through $W$ bosons or virtual squarks,

\be
\chcp_1\rightarrow (\LSP {W^+}^{(*)}, {\tilde{q}}^* q', \bar{q}
{\tilde{q'}}^*) \rightarrow \LSP \bar{q} q' \ ,
\ee
and so all six parameters enter the decay process. The lighter chargino
may also decay to LSPs through a virtual charged Higgs $H^{\pm}$, but
this diagram is suppressed by Higgs couplings and is negligible for all
but the most extreme choices of parameters. The heavier chargino may
decay through complicated cascade decays. However, when $\charginotwo$
production is kinematically accessible, the only information we will
use about $\charginotwo$ is its mass, which we will assume may be
measured through threshold scans.  The analysis will therefore be
independent of $\charginotwo$ branching fractions and other observables
dependent on the details of the $\charginotwo$ decay.

The chargino masses $\mchargino$ and $\mcharginotwo$ and the
right-handed cross section $\sigma_R$ depend only on the parameters
$\mu$, $M_2$, and $\tan\beta$, and these parameters may be used to
define regions with qualitatively different behavior. To understand
this, note first that, when $M_2 \gg |\mu|$ or $|\mu| \gg M_2$, the
following relations hold \cite{GH2}:

\be\label{thumb}
\mchargino \approx \min\{|\mu|, M_2\} \quad {\rm and} \quad
\mcharginotwo \approx \max\{|\mu|, M_2\} \ .
\ee
These relations are in fact approximately valid in most of the
available parameter space. The dependence of $\sigma_R$ on the
parameters is more complicated. In Fig.~\ref{fig:sigmar} we plot
contours of constant $\sigma_R$ for fixed $\tanb$ in the $(\mu, M_2)$
plane.  The dependence on $\tanb$ is fairly weak; we choose the
representative value $\tanb=4$ for illustration. Chargino production is
inaccessible for $\sqrt{s} = 500\gev$ in the hatched region, and the
cross-hatched region is excluded by the current experimental mass limit
$\mchargino > 45 \gev$ \cite{PDG,LEPgroups}. This leaves two bands, one
on each side of the $\mu=0$ axis. At the top of each band, where $M_2
\gg |\mu|$, the chargino is Higgsino-like, $\chargino \approx
\tilde{H}^{\pm}$, and we see that $\sigma_R$ is substantial.  However
as one moves into the region with $M_2 \alt |\mu|$, $\sigma_R$ quickly
drops.  This may be understood by noting that, because $\sqrt{s} \gg
\mz$, the $\gamma$ and $Z$ production diagrams may be replaced to a
good approximation by diagrams in which the $U(1)$ and $SU(2)$ gauge
bosons $B$ and $W^3$ are exchanged. However, the $e^-_R$ couples only
to $B$, and the $\tilde{W}^\pm$ couples only to $W^3$.  Thus, in the
region with $M_2 \alt |\mu|$, where the chargino is dominated by its
wino component and $\chargino \approx \tilde{W}^{\pm}$, the cross
section $\sigma_R$ is highly suppressed.

We are now in a position to define characteristic regions in the
parameter space.  These are shown for $\tanb=4$ in the $(\mu, M_2)$
plane in Fig.~\ref{fig:regions}. The hatched and cross-hatched regions
are as in Fig.~\ref{fig:sigmar}.  In the remaining area, we define the
following three regions, each of which includes a $\mu < 0$ part and a
corresponding $\mu>0$ part that is unlabeled:

\noindent Region 1: $\mchargino + \mcharginotwo < \sqrt{s}$. Here
$\chargino\chcflip_2$ production is possible, and so both chargino
masses can be measured.

\noindent Region 2 (shaded): $\mchargino + \mcharginotwo > \sqrt{s}$,
and $\sigma_R \alt 10 \fb$.

\noindent Region 3 (shaded): $\mchargino + \mcharginotwo > \sqrt{s}$,
and $\sigma_R \agt 50\fb$.

These three regions almost completely fill the region of  parameter
space in which chargino pair production is allowed at a 500 GeV
$e^+e^-$ collider, leaving only a small region in which the mixing is
large and the chargino $\charginotwo$ is just above threshold. In this
study, we will ignore this small gap.  In the two cases we will study
in detail, we will assume $\mchargino \approx 172 \gev$.  For this
value, the measurement of $\mchargino$ constrains the parameters to lie
on the dashed curves shown in Fig.~\ref{fig:regions}.  Then, if
$\charginotwo$ is not seen, $\sigma_R< 10 \fb$ or $\sigma_R> 70 \fb$
for $\tanb \ge 4$, and further, for $1 < \tanb < 4$, only small areas
of the $(\mu, M_2)$ plane lie outside regions 1--3.  For masses
$\mchargino$ nearer to threshold, the areas not covered by regions 1--3
are larger. However, this can be compensated by raising the collider
center-of-mass energy, which increases the size of region 1.

In region 3, if the ratio $M_1/M_2$ is fixed, $\chargino$ and $\LSP$
become increasingly degenerate as $M_2$ grows.  Charginos then decay to
invisible LSPs and very soft jets and leptons. It is therefore
difficult to choose a representative point in this region, as even the
identification of the chargino signal can be difficult in some areas.
More generally, if $M_1$ and $M_2$ are unrelated (and, of course,
independent of $\mu$), $\mchargino - \mLSP$ need not be small, even if
the chargino is Higgsino-like. Although it may then be possible to
verify SUSY relations in region 3, we will not consider this
possibility further.  However, we note that the MSSM makes a number of
nontrivial predictions for region 3.  Since $\chargino \approx
\tilde{H}^{\pm}$, the $\snu$ production diagram becomes negligible. The
production forward-backward asymmetry is thus approximately zero. In
addition, since the chargino is Higgsino-like, it decays predominantly
through a virtual $W$, and so the ratio of hadronic to leptonic decays
of the chargino should be equal to the corresponding ratio for $W$
bosons. These characteristic features should distinguish a chargino
candidate from new particles of other, non-supersymmetric origin.

\section{A Supersymmetry Test in the Mixed Region}
\label{sec:Mixed}

We now study a representative point in region 1 in detail. The
characteristic property of region 1 is that both chargino eigenstates
can be produced, and so both masses are measurable. Thus, in this
region, a promising approach will be to test the detailed form of the
chargino mass matrix.  In particular, notice that the matrix of
Eq.~(\ref{chamass}) contains, in addition to the new parameters $M_2$,
$\mu$, and $\tanb$, a dependence on the $W$ mass.  This is no accident.
The off-diagonal matrix elements of Eq.~(\ref{chamass}) result from the
$H \tilde{W} \tilde{H}$ vertex.  This is related by supersymmetry to
the $H W \partial H$ vertex, which is related by gauge invariance to
the term which gives mass to the $W$ through the Higgs mechanism. Thus,
verification that this parameter of Eq.~(\ref{chamass}) is indeed equal
to $\mw$ would be a quantitative test of supersymmetry.  This test is
formally independent of the neutralino sector and is therefore
applicable to models with gauge singlets.

We now investigate the extent to which we can realistically verify this
correspondence at the NLC. In this example, and for the rest of this
work, we will assume $\sqrt{s} = 500 \gev$.  We will present results
for integrated luminosities of 30 and 100 $\ifb$, corresponding roughly
to $\half$ to 2 years running at design luminosity.

For our case study, we choose the underlying supersymmetry parameters
to be

\be\label{region1pt}
(\mu, M_2, \tanb, M_1/M_2, \mslep, \msq) = (-195, 210, 4, 0.5, 400,
700) \ .
\ee
For these values, the MSSM gives

\be\label{region1quants}
\begin{array}{rcl}
\mchargino &=& 172 \gev \\
\mLSP &=& 105 \gev \\
\mcharginotwo &=& 255 \gev \\
(\phi_+,\phi_-) &=& (40.8^{\circ}, 59.5^{\circ}) \\
\sigma_R &=& 48 \fb \\
\sigma_L &=& 513 \fb \ .
\end{array}
\ee
For comparison, the QED $\mu^+ \mu^-$ production cross section is $397
\fb$.

To investigate the expected sensitivity to the form for the chargino
mass matrix, we generalize Eq.~(\ref{chamass}) to an arbitrary real
$2\times 2$ matrix, which we parametrize as

\be\label{chamasschi}
{\bf M'_{\chc}} = \left( \begin{array}{cc}
 M_2                    &\sqrt{2} \, \mwchi\sinbchi  \\
\sqrt{2} \, \mwchi\cosbchi &\mu                    \end{array}
\right) \ .
\ee
Without SUSY, the ratio of off-diagonal elements need not be the ratio
of vevs $\tanb \equiv \langle H^0_2 \rangle / \langle H^0_1 \rangle$,
and we have therefore replaced $\beta$ by $\beta^\chi$. As demanded by
gauge invariance, we also replace $\mz$ by $M_Z^\chi \equiv \mz
(\mwchi/\mw)$ in the neutralino mass matrix of Eq.~(\ref{neumass}).
\footnote{The resulting neutralino mass matrix is not the most general
allowed by gauge invariance.  The fully general neutralino mass matrix
will be considered briefly when neutralino events are considered in
Sec.~\ref{sec:Others}.}

We will investigate to what extent the NLC experiments may confirm the
SUSY relation $\mwchi = \mw$.  More explicitly, we have extended the
six-dimensional SUSY parameter space to a seven-dimensional parameter
space, and we will investigate how well experiments may reduce the
allowed region of this space to the supersymmetric subvolume in which
$\mwchi = \mw$. Formally, this is a simple task. The four parameters
entering Eq.~(\ref{chamasschi}) may be exchanged for the two masses and
two mixing angles, ($\mchargino$, $\mcharginotwo$, $\phi_+$, $\phi_-$).
By determining these four quantities from experiment, we can recover a
constraint on $\mwchi$.

To determine the chargino masses and mixing angles from experiment, we
will need to make assumptions about the decay properties of charginos.
In our analysis, we will assume that these properties are those of a
supersymmetric model at some point in parameter space, with the
exception that the new chargino and neutralino mass matrices are used.
Because we have not generalized the decay completely, this assumption
is a compromise, but we feel, a reasonable one --- it gives us a large
but well-defined space of possibilities to consider.  In addition, we
will see below, by explicitly scanning this space, that our results
depend only weakly on the decay parameters. The main dependences are
kinematic and would be expected in more general models of chargino
decays.  It is also worth noting that many of our assumptions may be
checked {\rm a posteriori}; for example, the assumption of a universal
left-handed slepton mass may be checked by observing the universality
of leptonic branching fractions in chargino decay.

The precision with which $\mchargino$ and $\mLSP$ can be determined was
studied by the JLC group~\cite{JLC}.  Using a method that depends on
kinematic arguments only, they found that, for an integrated luminosity
of 20 $\ifb$, these masses could be determined to approximately 2 GeV,
an uncertainty that is negligible for this study. The mass
$\mcharginotwo$ may be determined by scanning near $\chargino
\chcflip_2$ threshold. Although $\sigma (\epem \rarr \chargino
\chcflip_2)$ is suppressed by about an order of magnitude from mixing
angles, we will assume that an energy scan will be able to determine
$\mcharginotwo$ to a few GeV, and we will therefore also neglect this
uncertainty in the following analysis.

The crucial difficulty will be that of determining the two mixing
angles.  In principle, these can be extracted by measuring the
right-polarized differential cross section for $\chargino$ pair
production, which is completely determined by the $\chargino$ mass and
the two mixing angles. The right-polarized cross section $\sigma_R$,
though an order of magnitude smaller than $\sigma_L$, is still large
enough to yield a sufficient number of events for precision studies. In
particular, we will examine two quantities based on $\dsigr$: the total
cross section $\sigma_R$, and a truncated forward-backward asymmetry

\be\label{afbrdef}
\afbr \equiv \frac {\ds \sigma_R (0<\cos\theta<0.755)-
\sigma_R (-1<\cos\theta<0)} {\ds \sigma_R (-1<\cos\theta<0.755)} \ ,
\ee
where $\theta$ is defined as the angle between the $e^+$ beam and the
positive chargino $\chcp_1$. (The motivation for this peculiar
definition of $\afbr$ will be given below.) With $\mchargino$ known,
the values of $\sigma_R$ and $\afbr$ determine the variables ($\phi_+$,
$\phi_-$) and may therefore bound $\mwchi$. This strategy is appealing,
because we have seemingly eliminated all dependence on three of the
undetermined parameters of the theory: $M_1$, $\mslep$, and $\msq$.

Unfortunately, the analysis is not independent of these three
parameters when we consider what quantities are actually observable.
Cuts must be imposed to reduce standard model backgrounds.  In this
paper, we will rely on a standard set of cuts which have been
previously suggested to isolate the chargino pair production signal.
These  cuts select chargino events in which one chargino decays to an
isolated final state lepton, and the other decays {\it directly} to
hadrons. (Charginos may also decay indirectly to hadrons through $\tau$
leptons.) We will call such events ``$Y$ mode events,'' with the letter
``$Y$'' chosen to suggest the typical $2j+\ell$ topology of these
events. What is actually measured is not $\sigma_R$, but the $Y$ mode
partial cross section after cuts,

\be \label{sigydef}
\eta\sigy \equiv 2 \eta B_{\ell} B_h \sigma_R \ ,
\ee
where $\eta$ is the efficiency of the cuts for $Y$ events, $B_h$ is the
chargino branching ratio for direct hadronic decays, and $B_{\ell}$ is
the branching ratio for decays to a final-state lepton.  These
fractions both exclude decays to a $\tau$ which subsequently decays
hadronically.

Since the charginos decay very quickly, with typical widths of 1--100
keV,  the chargino direction and the asymmetry $\afbr$ cannot be
determined directly. We will measure $\afbr$ through its correlation to
$\afbhad$, the forward-backward asymmetry of the hadronic system in $Y$
events.  In principle, the experimentally observable quantities
$\afbhad$ and $\eta\sigy$  depend on the decay distributions, and thus
reintroduce dependence on the parameters $M_1$, $\mslep$, and $\msq$.
To understand the extent of this problem, we have performed Monte Carlo
simulations at a number of points in parameter space.  These points
have been chosen randomly, subject only to the constraints that they
give values of $\mchargino$, $\mLSP$, and $\mcharginotwo$ consistent
with those that would be measured in our case study. We will show below
that, in the resulting subvolume of parameter space, the experimental
observables turn out to be rather insensitive to $M_1$, $\mslep$, and
$\msq$, and therefore the virtues of our strategy in fact remain.

To simulate chargino events, we used the parton level Monte Carlo event
generator of Ref.~\cite{LEPIIstudy}.  This generator includes the spin
correlations between production and decay processes. To simulate
hadronization and detector effects, the final state partons were
smeared with detector parameters as chosen in the JLC study \cite{JLC}:

\be
\frac{\sigma^{\rm had}_E}{E} = \frac{40 \%}{\sqrt{E}}
\quad {\rm and} \quad
\frac{\sigma^{\rm lepton}_E}{E} = \frac{15 \%}{\sqrt{E}} \ ,
\ee
where $E$ is in GeV.

The $Y$ chargino events were selected by first using a system of cuts
presented in Ref.~\cite{Fujii}.  These cuts are designed for charginos
that decay through off-shell $W$ bosons, and include the following:

\noindent (a) $|\cos \theta_i | < 0.9$ for every final state parton,
where $\theta_i$ is the polar angle of parton $i$ with respect to the
$e^+$ beam axis.

\noindent (b) $E_{\ell} > 5 \gev$, $\theta_{q\ell} > 30^\circ$, that
is, there must be an energetic $e$ or $\mu$ with no hadronic activity
within a cone of half angle $30^{\circ}$.

\noindent (c) $20 \gev < E_{\rm visible} < \sqrt{s}-100 \gev$.

\noindent (d) $\theta_{\rm acoplanarity} < 150^\circ$.

\noindent (e) $m_{\rm had} < 68 \gev$, $E_{\rm had} < \sqrt{s}-100
\gev$, where $m_{\rm had}$ and $E_{\rm had}$ are the mass and energy of
the hadronic system.

\noindent (f) $|m_{\ell\nu} - \mw| > 10 \gev$, where the $\nu$ momentum
is taken to be equal to the missing momentum.

\noindent (g) $-Q_{\ell} \cos\theta_{\rm had}, Q_{\ell}
\cos\theta_{\ell} < \cos 41^{\circ} = 0.755$, where $Q_{\ell}$ is the
charge of the isolated lepton, and $\theta_i$ is as defined in cut (a).

\noindent These cuts isolate chargino events that have hadrons and an
isolated lepton in the final state.  We would like to isolate $Y$
events, and we therefore need to eliminate events in which the hadronic
system results from charginos decaying through $\tau$ leptons.  This
may be done by imposing the additional requirement that the mass of the
hadronic system $m_{\rm had}$ be greater than $m_{\tau}$. As was shown
in Ref.~\cite{LEPIIstudy}, $Y$ events very rarely have low $m_{\rm
had}$ at LEP II energies, and we have verified that this is also true
for NLC energies.  We will therefore simply assume that this additional
cut on $m_{\rm had}$ cleanly isolates the $Y$ mode events.

Cuts (c) and (d) are efficient for supersymmetric signals because of
the large momentum and energy that are carried off by the unobserved
massive LSPs. Cuts (e)--(g) reduce the dominant standard model
background, $W$ pair production. In particular, cut (g) is designed to
remove the large forward peak of $WW$ events. Because the hadronic
system's polar angle distribution is truncated by cut (g), we choose
$\afbr$, as defined in Eq.~(\ref{afbrdef}), as the theoretical quantity
with which we expect $\afbhad$ to be well-correlated. Since $W$ pair
production results primarily from $e^-_L e^+$ annihilation, the use of
these cuts in conjunction with a very highly right-polarized $e^-$ beam
results in a negligible background rate.  The analysis of
Ref.~\cite{Fujii} included $t\bar{t}$ events with a top quark mass of
150 GeV and found negligible background from this source.

We caution the reader that the cuts (a)--(g) above have been designed
to separate the chargino signal from standard model backgrounds, but
have not been optimized to discriminate between $\chcp_1 \chcm_1$
production and other SUSY sources of $Y$ events. In principle, these
could include $\chc_1\chcflip_2$ and $\chcp_2 \chcm_2$ production, as
well as the production of neutralino pairs $\chn_i \chn_j$.  Ignoring
effects of resolution smearing, the neutralino events will be
backgrounds to $Y$ events only when a heavy neutralino decays into a
chargino and a $W$ boson, which then decays leptonically to provide the
single isolated lepton. While we have not simulated these events, we do
not expect neutralinos to be a severe background because their
production cross sections are generally small, and further, their
decays to $h \LSP$ and $Z \LSP$ are usually favored by phase space and
therefore dominate. For the point that we are studying, the masses of
the heavy neutralinos are $m_{\chn_2} = 169 \gev$, $m_{\chn_3} = 211
\gev$, and $m_{\chn_4} = 253 \gev$.  The decay $\chn_4 \rightarrow
W\chargino$ is barely open, and the production of heavy chargino pairs
is kinematically forbidden. Thus, $\chc_1\chcflip_2$ production, with
$\chc_2\rightarrow W^{\pm}\LSP \rightarrow \ell^{\pm}\nu \LSP$ is the
main SUSY contamination in the present case study. This background is
restricted by phase space and mixing angles and can be eliminated
entirely by running below the $\chc_1\chcflip_2$ production threshold.

Throughout this study, we have assumed 100\% beam polarization in our
simulations.  In the present case, however, because $\sigma_L$ is an
order of magnitude larger than $\sigma_R$, the left-handed
contamination of the right-handed beam could be substantial if the beam
polarization is not nearly 100\%.  If beam polarization near 100\% is
unobtainable, the $e^-_R$ signal may be determined by first measuring
the $e^-_L$ signal to high accuracy, and then subtracting the
left-handed contamination from the right-polarized $e^-$ beam's signal.
For a beam polarization of 95\%, these errors will not be large, and we
have not included the statistical errors resulting from such a
subtraction. It is clear, however, that highly polarized beams play a
critical role in reducing such errors.

We now determine the correlation of $\afbr$ with $\afbhad$ through
Monte Carlo simulations.  A description of our method and the relevant
formulae are contained in the appendix. We sample random points in the
seven dimensional parameter space, with only the restriction that
$\mchargino$, $\mLSP$, and $\mcharginotwo$ are each within 2 GeV of
their values in Eq.~(\ref{region1quants}).  For each set of parameters,
we calculate $\afbr$ from explicit analytical formulae and determine
$\afbhad$ through Monte Carlo simulation.  The results for 38
simulations are plotted in Fig.~\ref{fig:afb}. A simple linear fit
yields $\afbhad = 0.717 \afbr + 0.042 \pm 0.036$, where $\amctot =
0.036$ is the $\onesigma$ deviation in $\afbhad$ for a fixed $\afbr$.
The best fit is given by the solid line in Fig.~\ref{fig:afb}, and the
$\onesigma$ deviations are shown by the dashed lines.

However, this quoted error overestimates the deviation from perfect
correlation between $\afbr$ and $\afbhad$, because each point in
Fig.~\ref{fig:afb} was computed from a finite sample of Monte Carlo
events and therefore contains a non-negligible  statistical
fluctuation. The average effective number of Monte Carlo events for the
simulations was $N_{\rm MC} \approx 1400$. Using the formulae contained
in the appendix, we find that the Monte Carlo statistical error is
$\amcstat = 0.026$; when this is removed, the systematic error in
assuming perfect correlation is found to be $\asys = 0.025$. The
correlation between $\afbr$ and $\afbhad$ is high --- the chargino rest
frames are slightly boosted, and the decay distributions are
sufficiently similar for all sampled values of the underlying
parameters that $\afbhad$ is highly insensitive to the decay process
and is well-determined for a fixed $\afbr$.  (If the beam energy is
slightly reduced to run below the $\chc_1\chcflip_2$ threshold, the
charginos will be less boosted.  However, we do not expect the
correlation between $\afbr$ and $\afbhad$ to deteriorate much, since,
even in the present case with $\sqrt{s} = 500 \gev$ and only slightly
boosted charginos, the correlation is high.)

To determine the bounds that may be placed on $\afbr$ experimentally,
we must add the experimental statistical error to $\asys$. For our
representative point, a Monte Carlo simulation gives

\be
\begin{array}{rcl}
\afbhad &=& -0.233 \\
\eta    &=& 35.5\% \\
N_{\rm exp}   &=& 6.0 {\cal L}_R \ ,
\end{array}
\ee
where $N_{\rm exp}$ is the number of $Y$ events surviving the cuts, and
${\cal L}_R$ is the right-handed integrated luminosity in $\ifb$. The
total experimental uncertainties for two values of right-polarized
integrated luminosity are found to be

\be
{\cal L}_R = 30 \, (100)\, \ifb \Longrightarrow
\afbr  = -0.37 \pm 0.107 \, (0.065) \ .
\ee

The efficiency $\eta$ also depends on the decay process. We determine
$\eta$ by finding its range in the subvolume of parameter space in
which the three masses and $\afbr$ are within the experimental bounds
of their underlying values. Each simulation gives a point in the
$(\afbr, \eta)$ plane, and the distribution of points is plotted in
Fig.~\ref{fig:eta}. A linear fit gives $\eta = - 6.48 \afbr + 34.35 \pm
1.07\%$, where $\emctot = 1.07\%$ is the $\onesigma$ deviation in
$\eta$ for a fixed $\afbr$. As in the previous figure, the best linear
fit is given by the solid line, and the dashed lines give the $1\sigma$
deviations.  We see that there is a dependence on $\afbr$ --- in cases
in which chargino production is forward peaked, cut (g) lowers the
efficiency.  However, since we have already bounded $\afbr$ in the
analysis above, we may use this measurement to restrict the range of
$\eta$.  To determine the systematic error, we remove the Monte Carlo
statistical error from $\emctot$.  Following the analysis of the
appendix, we find that $\emcstat = 0.77\%$ and $\esys = 0.75\%$, and,
including experimental statistical errors, we find

\be
{\cal L}_R = 30 \, (100)\, \ifb \Longrightarrow
\frac{\Delta\sigy}{\sigy} = 8.0 \, (4.7) \, \% \ .
\ee

To convert a measurement of $\sigy$ into a measurement of  $\sigma_R$,
we must also take into  account the uncertainty in the branching ratios
$B_{\ell}$ and $B_h$.  These again depend on the parameters of the
chargino decay matrix elements and, in particular, on the masses
$\mslep$ and $\msq$.  We have varied these masses to permit as a large
a variation in $\sigma_R$ as possible.  However, the measurements of
$\mchargino$, $\mcharginotwo$, $\afbr$, and $\sigy$ constrain the
allowed parameter ranges to regions where $\chcp_1$ and $\chcm_1$ have
substantial Higgsino components.  Recall that $B_{\ell}$ and $B_h$ take
fixed values (equal to those for the $W$) in the Higgsino limit.  These
facts and the bounds $\mslep , \msq > 250 \gev$ constrain the $Y$ mode
branching fraction to the region in which $29\% < 2 B_{\ell} B_h <
36\%$. Thus, the $\sigy$ contours are rather insensitive to variations
in the sfermion mass parameters.

The measurements of $\afbr$ and $\sigma_R$ constrain the $(\phi_+,
\phi_-)$ plane to the shaded regions in Fig.~\ref{fig:phiplane}.  The
lightly (heavily) shaded region is the allowed region for ${\cal L}_R =
30 \, (100) \, \ifb$. Contours of constant $\mwchi$ are also plotted in
GeV, with the SUSY contour $\mwchi = \mw$ given by the dotted curves.
The contours of constant $\sigma_R$ that bound the allowed region run
roughly northwest to southeast; contours of constant $\afbr$ run
roughly southwest to northeast.  The indicated boundaries correspond to
1$\sigma$ deviations in each quantity.

Given the chargino masses of this case study, the theoretically
possible range of $\mwchi$ is

\be
0 \le \mwchi \le \left(\frac{\mchargino^2 + \mcharginotwo ^2}{2}
\right) ^{\frac{1}{2}} = 218 \gev \ .
\ee
In the allowed region for ${\cal L}_R = 100 \ifb$,

\be
60 \gev < \mwchi < 105 \gev \ .
\ee
The measurement of $\mwchi$, therefore, provides a quantitative
confirmation of SUSY.

As an aside, we note that our analysis simultaneously bounds the
parameters $\mu$, $M_2$, and $\tanbchi$.  In the heavily shaded region,
the allowed ranges for these parameters are

\be
\begin{array}{rcccl}
-204 \gev &<&\mu &<& -183 \gev \\
199 \gev &<& M_2 &<& 217 \gev \\
2.4 &<& \tanbchi &.&
\end{array}
\ee
If one is led by the bounds on $\mwchi$ (or other considerations) to
view SUSY and the MSSM as confirmed, one might then consider only the
contour $\mwchi = \mw$ within the allowed region.  One would also be
led to identify $\tanbchi$ with the ratio of Higgs scalar vevs, and so
we will replace $\beta^\chi$ with $\beta$. On the contour $\mwchi =
\mw$, the bounds on the SUSY parameters are extremely strong:

\be\label{3mainbounds}
\begin{array}{rcccl}
-196 \gev &<& \mu &<& -193 \gev \\
208 \gev &<& M_2 &<& 211 \gev \\
3.9 &<& \tanb &<& 4.1 \ .
\end{array}
\ee
These bounds are so strong that it is likely that the uncertainties in
chargino masses will be a significant source of uncertainty.  (Recall
that, while the uncertainties in chargino masses were included in the
determination of systematic errors, the parameter bounds are determined
{}from Fig.~\ref{fig:phiplane}, in which the chargino masses are
fixed.) Nevertheless, it is clear that the discovery of both chargino
mass eigenstates will allow one to place tight bounds on these three
central SUSY parameters. In particular, the bound on $\tanb$ would be
one of the most stringent and model-independent; the difficulty of
determining $\tanb$ from the Higgs scalar sector is explained in
Ref.~\cite{Janot}. Given the bounds of Eq.~(\ref{3mainbounds}), other
SUSY parameters may be restricted by additional measurements. For
example, $\mLSP$ may be used to determine $M_1$, and $\sigma_L$ may be
used to find $\mslep$. Such determinations may help lead us to an
understanding of the SUSY breaking mechanism and other aspects of
higher theories.

We have now completed the case study for our chosen representative
point.  We conclude this section with comments concerning the power of
this analysis for other points in region 1. If one moves from the point
given in Eq.~(\ref{region1pt}) toward region 3, the results of the
analysis become stronger for two reasons. First, $\sigma_R$ increases,
and the experimental statistical errors decrease.  Second, as a direct
consequence of electroweak gauge invariance, such large values of
$\sigma_R$ can only be achieved for Higgsino-like $\chargino$, even in
the generalized (seven-dimensional) parameter space where one lets
$\mwchi$ vary.  This implies that chargino decay is dominated by the
$W$ diagram, and the sensitivity to the decay process parameters
becomes even weaker than in our case study.  In particular, the
systematic errors related to determining $\afbr$ and $\eta$ become
smaller, and the branching ratios $B_{\ell}$ and $B_h$ take their $W$
decay values.

If one moves in the opposite direction toward region 2, the number of
right-polarized events deteriorates rapidly.  In addition, $\chargino$
may be gaugino-like, and the branching fractions therefore depend more
strongly on decay parameters, leading to a larger uncertainty in the
determination of $\sigma_R$ from $\sigy$.  These problems can
potentially be remedied by changing the analysis method.  Since a
highly right-polarized $e^-$ beam leads to a very small level of
background, it may be possible to use a  looser system of cuts, and to
measure the hadronic and leptonic branching fractions directly. The
analysis in the gaugino-like portion of region 1 would then be limited
only by statistics and systematic errors in the determination of
$\afbr$ and $\eta$, and the statistical uncertainties in the
measurements of the branching fractions for chargino decays.

Finally, having considered variations of the Higgsino-gaugino content
of $\chargino$, one might consider variations orthogonal to these in
the plane of Fig.~\ref{fig:regions}, namely, variations in
$\mchargino$.  If $\chargino$ is heavier, the chargino rest frame is
less boosted relative to the lab frame.  The decay process will then
have a bigger effect on the correlation of $\afbhad$ with $\afbr$, and
$\asys$ will increase.  However, we have already considered a case with
a fairly heavy $\chargino$, and we see that the charginos need not be
highly relativistic for $\asys$ to be small.  In the opposite limit of
lighter $\chargino$, the chargino rest frame is more boosted relative
to the lab frame, decay effects become less important, and the results
of our analysis can be expected to improve.

\section{A Supersymmetry Test in the Gaugino Region}
\label{sec:Gaugino}

In the previous section, we considered the case in which both charginos
were discovered, and found that the SUSY constraint on the chargino
mass matrix could be verified to fairly high precision.  In this
section, we examine region 2, in which only one chargino is seen and
its production cross section section from $e^-_R$ is small.  Here we
must rely on the chargino pair production cross section from $e^-_L$,
which introduces a strong dependence on $\msnu$ from the second diagram
in Fig.~\ref{fig:productiondiagrams}.  Fortunately, there is an
important compensating simplification: in this region, the charginos
are very nearly pure gauginos, and, in fact, it is a good approximation
to neglect the deviations of $\cos\phi_\pm$ from 1.  In this limit, the
coupling constant of the $e^{\mp}\snu\chargino$ vertex is related by
supersymmetry to the $e^{\mp}\nu W^{\pm}$ coupling constant $g$.
Verification that this coupling constant is indeed equal to $g$ would
be a quantitative test of supersymmetry.

For our case study in region 2, we take the underlying supersymmetry
parameters to be

\be
(\mu, M_2, \tanb, M_1/M_2, \mslep, \msq) = (-500, 170, 4, 0.5, 400,
700) \ .
\ee
For these values, the MSSM gives

\be
\begin{array}{rcl}
\mchargino &=& 172 \gev \\
\mLSP &=& 86 \gev \\
\mcharginotwo &=& 512 \gev \\
(\phi_+,\phi_-) &=& (1.2^{\circ}, 12.8^{\circ}) \\
\sigma_R &=& 0.15 \fb \\
\sigma_L &=& 612 \fb \ .
\end{array}
\ee
For the point we have chosen (and for a significant part of region 2),
the two-body chargino decay $\chargino \rightarrow W^{\pm} \LSP$ is
open. The branching fractions $B_{\ell}$ and $B_h$ are then fixed to
their values in $W$ decay, unless $|\mu|$ is very large, a possibility
discussed at the end of this section. The case in which on-shell $W$
decays are not allowed will also be discussed briefly at that point.

To investigate the sensitivity of experiments to the value of the
$e^{\mp}\snu\chargino$ coupling, we generalize this coupling from its
SUSY value $g {\bf V}_{11}$ to $\gchi {\bf V}_{11}$.  We then test the
SUSY relation $\gchi=g$. The differential cross section $\dsigl$ is
then a function of $(\mchargino, \phi_+, \phi_-, \msnu, \gchi)$, but
because $\phi_+, \phi_- \approx 0$ and we can measure $\mchargino$, we
have only two unknowns.  These may be constrained with two quantities
formed from $\dsigl$, which we choose to be $\sigma_L$ and

\be
\afbl \equiv \frac {\ds \sigma_L (0<\cos\theta<0.707)-\sigma_L
(-1<\cos\theta<0)} {\ds \sigma_L (-1<\cos\theta<0.707)} \ .
\ee

It is important to note that the parameters $\gchi$ and $\msnu$ enter
$\dsigl$ only through the $\snu$ diagram amplitude, which has the form

\be
A_{\snu} \sim \frac{|\gchi {\bf V}_{11}|^2}{t-\msnu^2} \ .
\ee
Thus, for very large values of $\msnu$, only the ratio $\gchi/\msnu$
can be determined.  However, we will see that even for the rather large
value of $\msnu$ that we have chosen, the two parameters $\gchi$ and
$\msnu$ can be distinguished.  In general, these parameters can be
bounded independently when $\msnu$ is comparable to the collider
center-of-mass energy (though still possibly above the pair-production
threshold).

We follow the procedure of the previous section, with the exception of
using the cuts of Ref.~\cite{Grvz2}, which are appropriate for
charginos decaying through on-shell $W$ bosons. These include the
following:

\noindent (a) $E_{\ell} > 5 \gev$, $\theta_{q\ell} > 60^\circ$.

\noindent (b) $\mpt > 35 \gev$.

\noindent (c) $\theta_{\rm acoplanarity} < 150^\circ$.

\noindent (d) $|m_{\ell\nu_{\rm ISR}} - \mw| > 10 \gev$, where
$\nu_{\rm ISR}$ is defined to be the massless particle which, along
with an initial state radiated photon in the $\pm \hat{z}$ direction,
makes up the missing momentum.

\noindent (e) $\theta_{\rm sphericity} < 45^\circ$, which we
approximate in the Monte Carlo simulation by demanding $-Q_{\ell}
\cos\theta_{\rm had}$, $Q_{\ell} \cos\theta_{\ell} < \cos 45^{\circ} =
0.707$.

\noindent This system of cuts isolates chargino events containing
hadrons and an isolated lepton.  Again, the subset of these events that
are $Y$ mode events may be cleanly separated by demanding that $m_{\rm
had}$ be significantly larger than $m_{\tau}$. After these cuts, the
$WW$ background is reduced to roughly 25 fb for an $e^-_L$ beam, which
is approximately the size of the signal after cuts.  We will assume
that the $WW$ background is well-understood and may be subtracted up to
statistical fluctuations.  As the $WW$ background is strongly
forward-peaked, we will also assume in computing statistical errors
that it contributes completely to the set of events with $\cos\theta >
0$.  The $t\bar{t}$ background, computed with $m_t = 140 \gev$, is
again negligible.  In the gaugino region, the other SUSY signals do not
provide a significant background to $Y$ events, because the only
kinematically accessible SUSY backgrounds are $\chn_2\LSP$ and
$\chn_2\chn_2$, with $m_{\chn_2} \approx \mchargino$. The neutralinos
$\chn_2$ then decay to LSPs and an even number of leptons, and the
number of events with one isolated lepton is highly suppressed.

To determine the correlation between $\afbhad$ and $\afbl$, we perform
Monte Carlo simulations at a number of randomly chosen points in the
seven-dimensional parameter space ($\mu$, $M_2$, $\tanb$, $M_1/M_2$,
$\mslep$, $\msq$, $\gchi$), subject to the constraints that
$\mchargino$ and $\mLSP$ are within 2 GeV of their measured values and
$\sigma_R < 1 \fb$. Again the experimental observable $\afbhad$ is
determined to be an excellent estimator of $\afbl$, with $\asys =
0.034$. A Monte Carlo simulation at our representative point gives

\be
\begin{array}{rcl}
\afbhad &=& 0.034 \\
\eta    &=& 11.9\% \\
N_{\rm exp}   &=& 25.8 {\cal L}_L \ ,
\end{array}
\ee
where ${\cal L}_L$ is the left-handed integrated luminosity in $\ifb$.
We now calculate the uncertainties in determining $\afbl$ and $\sigy$
using the equations found in the appendix, this time including also the
errors arising from a substantial number of background events $N_{\rm
back} \approx N_{\rm exp}$.  We find that for two values of
left-polarized integrated luminosity,

\be
{\cal L}_L = 30 \, (100)\, \ifb \Longrightarrow
\afbl  = 0.20 \pm 0.067 \, (0.048) \ .
\ee

As in the previous case, the efficiency $\eta$ is found to be highly
constrained by the measurements of $\mchargino$, $\mLSP$, $\sigma_L$,
and $\afbl$, and the resulting systematic error is $\esys = 0.55\%$.
Including experimental statistical errors and those resulting from
background subtraction, we find

\be
{\cal L}_L = 30 \, (100)\, \ifb \Longrightarrow
\frac{\Delta\sigy}{\sigy} = 7.2 \, (5.6) \, \% \ .
\ee

For ${\cal L}_L = 30$ and 100 $\ifb$ these measurements constrain the
allowed region of the $(\msnu,\gchi)$ plane to the shaded areas shown
in Fig.~\ref{fig:msnugchiplane}. Because the charginos decay through
on-shell $W$ bosons, in contrast to the region 1 analysis, $B_{\ell}$
and $B_h$ are fixed at their values in $W$ decay, and thus the contours
for $\sigma_L$ inferred from $\sigy$ are independent of sfermion
masses. For ${\cal L}_L = 100 \ifb$, if $\msnu< 250 \gev$ is excluded
by the non-observation of any other threshold for heavy particle
production, the allowed region is only the largest of the three shaded
regions in Fig.~\ref{fig:msnugchiplane}b. For this region, we find the
constraint

\be
0.85g\le \gchi \le 1.3g \ .
\ee
Such a result would be an important quantitative confirmation of SUSY.

Fig.~\ref{fig:msnugchiplane} also illustrates a number of other
interesting features.  It is clear from Fig.~\ref{fig:msnugchiplane}
that, without assuming SUSY, the analysis above has simultaneously
bounded the mass $\msnu$ of a  $t$-channel resonance, a useful result
for future particle searches.  If, on the other hand, we assume the
validity of SUSY, then we are restricted to the dotted line at
$\gchi/g=1$, and the $\afbl$ measurement {\it alone} restricts $\msnu$.
Alternatively, the $\sigma_L$ measurement {\it alone} restricts $\msnu$
to two different ranges, of which one can be immediately excluded.
Finally, as expected from earlier comments, this analysis is
significantly weakened if $\msnu$ is large. For large $\msnu$, the
contours of Fig.~\ref{fig:msnugchiplane} approach contours of constant
$\gchi/\msnu$, and only the ratio $\gchi/\msnu$ can be determined.  On
the other hand, if it is possible to measure $\msnu$ independently, for
example, from $e^{\mp} \snu \chargino$ production, then the bounds on
$\gchi/g$ can be significantly improved.

In the example above, we have considered a point for which chargino
decays through on-shell $W$ bosons are allowed.  This choice was
motivated by two considerations.  First, in region 1, we considered a
point for which only off-shell $W$ decays were possible, and
appropriate cuts were used.  Our choice in region 2 illustrates that
tests of SUSY are also possible when cuts appropriate to on-shell $W$
decays must be used. Second, the scenario in which on-shell $W$ decays
are possible becomes more and more typical as the chargino mass rises,
and the analysis presented is thus generalizable to higher chargino
masses and beam energies. It is easy, however, to find points in region
2 where the chargino cannot decay to an on-shell $W$. For example, if
one assumes the GUT relation $M_2 = 2 M_1$, on-shell $W$ decays are
excluded for $\mchargino \alt 160 \gev$.  In this case, we must use the
cuts presented in Sec.~\ref{sec:Mixed}.  In addition, chargino decays
through virtual sfermions are not negligible, and one must consider the
dependences of the branching ratios on sfermion masses. Such
dependences will introduce systematic errors that may considerably
weaken our results. However, as in the case of the gaugino portion of
region 1, if these branching ratios can be measured, the systematic
errors in their determination may be greatly reduced. In contrast to
the region 1 case, the $e^-_L$ beam, with its accompanying $WW$
background, must be used. However, because $WW$ events do not usually
have $\mpt$ without isolated leptons, they are likely to be a small
background to purely hadronic chargino events.  Although further study
is required, it again seems probable that the $Y$ mode branching
fraction can be measured directly, and, with these modifications, the
previous analysis may be applied to region 2 scenarios in which only
off-shell $W$ decays are allowed.

It is also true that in the very far gaugino region with $|\mu| \gg
M_2$, where $\chargino \approx \tilde{W}^{\pm}$ and $\LSP \approx
\tilde{B}$, the $W$ decay diagram is suppressed by mixing angles, and,
even when decays through on-shell $W$ bosons are kinematically allowed,
virtual sfermion diagrams may be important. This requires that
$\chargino$ and $\LSP$ be very nearly pure gauginos, however, and this
occurs only for $|\mu| \agt 1 \tev$, a condition that is disfavored by
fine-tuning constraints.

\section{Sfermions and Neutralinos}
\label{sec:Others}

Up to this point, we have considered only precision SUSY tests from
studies of the properties of charginos.  Other sparticles may be
produced at NLC energies, however, and we now examine the possibility
of testing SUSY through the properties of sfermions and neutralinos.
The discussion will be limited to brief remarks and, in contrast to the
previous sections, no attempt will be made to perform detailed studies.

We first investigate the possibility of identifying a few
newly-discovered scalars as sfermions.  We are most interested in the
scenario in which these scalars provide the first opportunity for
precision tests of SUSY, and we therefore consider the case in which
these scalars are lighter than charginos. In contrast to the previous
sections, we will not impose any constraints on intergenerational
slepton and squark mass degeneracies. However, if the problem is
considered in full generality, it is complicated by many arbitrary
parameters associated with sfermion intergenerational mixing. Simply to
make the problem tractable, we will assume that intergenerational
mixing is absent. We will also assume that left-right mixings may be
neglected, with the understanding that the discussion that follows may
not be applicable to the sfermions of the third generation.  Probes of
the left-right mixing of scalar taus have recently been discussed by
Nojiri \cite{Nojiri}.

With these assumptions, the properties of these sfermions are
completely specified by their quantum numbers and their masses. The
only category (III) tests involving sfermion properties are therefore
verifications of mass relations. Given the assumptions above, the
masses of sfermions are

\be
\begin{array}{rcl}
m_{\tilde{u}_L}^2 &=&
m_{\tilde{Q}}^2+m_u^2+\mz^2(\half-\twothird\sinsqw )\cos 2\beta  \\
m_{\tilde{d}_L}^2 &=&
m_{\tilde{Q}}^2+m_d^2+\mz^2(-\half+\third\sinsqw )\cos 2\beta \\
m_{\tilde{u}_R}^2 &=&
m_{\tilde{U}}^2+m_u^2+\mz^2(\twothird\sinsqw )\cos 2\beta \\
m_{\tilde{d}_R}^2 &=&
m_{\tilde{D}}^2+m_d^2+\mz^2(-\third\sinsqw )\cos 2\beta \\
m_{\tilde{e}_L}^2 &=&
m_{\tilde{L}}^2+m_e^2+\mz^2(-\half+\sinsqw )\cos 2\beta \\
m_{\tilde{\nu}_L}^2 &=& m_{\tilde{L}}^2+\half\mz^2\cos 2\beta  \\
m_{\tilde{e}_R}^2 &=&
m_{\tilde{E}}^2+m_e^2+\mz^2(-\sinsqw )\cos 2\beta \ ,
\end{array}
\ee
where $m_{\tilde{Q}}$, $m_{\tilde{U}}$, $m_{\tilde{D}}$,
$m_{\tilde{L}}$, and $m_{\tilde{E}}$ are soft SUSY breaking scalar
masses.  Similar relations hold for second and third generation
sfermions.  With additional relations from grand unification, there are
a number of relations among these scalar masses \cite{Martin}. However,
we will continue to eschew assumptions that are not phenomenologically
motivated.  Without GUT assumptions, the right-handed masses are
unrelated to the other masses, and the left-handed masses are related
only by

\be\label{massdiff}
\begin{array}{rcl}
m_{\tilde{e}_L}^2 - m_{\tilde{\nu}_L}^2 &=& -\mw^2 \cos 2\beta  \\
m_{\tilde{d}_L}^2 - m_{\tilde{u}_L}^2 &=&-\mw^2 \cos 2\beta  \ ,
\end{array}
\ee
where we have omitted the small fermion mass terms.  For $\tanb > 1$,
these mass differences are positive, but we will consider all possible
values of $\tanb$ below. The relations of Eq.~(\ref{massdiff}) are
quantitative predictions of SUSY that we may try to test.

Unfortunately, if the newly discovered scalars are sleptons, it will be
impossible to test these relations, because the sneutrinos will decay
invisibly through $\snu_L \rarr \nu \LSP$. We are assuming that
charginos are heavier than sneutrinos, and so the decay $\tilde{\nu}
\rarr e_L \chcp_1$ is excluded. Also, even if $\tanb < 1$ and
$m_{\tilde{\nu}_L} > m_{\tilde{e}_L}$, the experimental lower bound
$m_{\tilde{e}_L} > 45 \gev$ \cite{PDG,LEPgroups} implies
$m_{\tilde{\nu}_L} - m_{\tilde{e}_L} < 47 \gev$, and the decay
$\tilde{\nu}_L \rarr \tilde{e}_L W^+$ is also greatly suppressed. The
masses of sleptons are therefore highly unlikely to provide the {\it
first} category (III) verifications of SUSY. Of course, if sneutrinos
are heavier than charginos, precise verifications of slepton mass
relations could be used to supplement measurements of chargino
properties. It should also be noted that other properties of slepton
{\it events} may provide additional precision measurements in the
gaugino region. It may be possible, for example, to measure some
neutralino properties through the $t$-channel $\chn_i$ exchange
diagrams for charged slepton pair production \cite{Fujii}.

On the other hand, if the scalars are squarks, both left-handed species
will decay visibly.  A previous study of squark mass determination
found that at the NLC, in most regions of parameter space, squark
masses can be measured to approximately 2 GeV with an integrated
luminosity of $10 \ifb$, even in scenarios with cascade decays
\cite{squark}.  This study also found that left-handed squarks can be
effectively separated from right-handed squarks using beam
polarization. It may be difficult to properly assign flavors to the
different squark mass thresholds, however, especially if these
thresholds are not well-separated. Let us first suppose that the masses
of only two left-handed squarks are determined. To verify SUSY
quantitatively, one must assume that the squarks are in the same
generation, and must also independently determine $\tanb$ from the
Higgs scalar sector.  This is by no means always possible, and most
likely requires, for example, that $m_{A^0} \alt 300 \gev$ so that a
heavy Higgs boson is kinematically accessible \cite{Janot}. Even if all
of these measurements can be made, the precision of the test is not
high. For example, if $\msq > 200 \gev$, the mass difference is
$|m_{\tilde{u}_L} - m_{\tilde{d}_L}| < 15 \gev$, and so in the best
case scenario where $\tanb$ is determined exactly, the squark mass
relation can be verified to approximately 20\%. If it is not possible
to measure $\tanb$ from the Higgs boson sector, a precision test of
squark mass relations is only possible if one measures four left-handed
squark masses.  One can then check that there exists some flavor
assignment consistent with

\be\label{massdiff2}
m_{\tilde{d}_L}^2 - m_{\tilde{u}_L}^2 \approx
m_{\tilde{s}_L}^2 - m_{\tilde{c}_L}^2 \equiv \Delta (m^2) \ ,
\ee
where $|\Delta (m^2)| \le \mw^2$.

The possibility of making the first quantitative tests of SUSY from
sfermion properties is therefore not very promising.  In the case of
sleptons, the prospects are bleak, while in the case of squarks, even
after assuming that intergenerational mixing is absent, precision tests
are complicated by difficulties in flavor determination and rely on
many MSSM scalars being kinematically accessible.  However, the
sfermion sector provides a number of opportunities for disproving the
MSSM and SUSY. For example, if sneutrino decay is observed, one of our
assumptions must be invalid. Also, the relations of
Eq.~(\ref{massdiff}) are valid not just for the MSSM, but are extremely
general predictions of SUSY.  If they are found to be violated, not
only will the MSSM be excluded, but almost all supersymmetric models
will be strongly disfavored. On the other hand, of SUSY is favored by
experiment, measurements of the squark and slepton masses will give
important information about the flavor dependence of the SUSY breaking
mechanism.

Neutralinos are natural candidates for precision SUSY tests, because,
with the assumption that the lightest neutralino $\LSP$ is the LSP, all
sparticle event observables depend, at least formally, on the
parameters that determine neutralino properties. In addition,
neutralinos are light in many models, and, in fact, throughout
parameter space, if charginos are produced, $\chn_1 \chn_2$ production
is kinematically possible.

One might hope to follow the procedure in Sec.~\ref{sec:Mixed} by
generalizing the neutralino mass matrix. If we relax SUSY, the most
general form of Eq.~(\ref{neumass}) consistent with gauge invariance is

\be\label{neumasschi}
{\bf M'_{\chn}} =
\left( \begin{array}{cc}
\begin{array}{cc} M_1 & 0    \\ 0   & M_2 \end{array}
& {\cal M} \\
{\cal M}^T
& \begin{array}{cc} 0   & -\mu \\ -\mu& 0 \end{array}
\end{array}
\right) \ ,
\ee
where ${\cal M}$ is an arbitrary $2\times 2$ matrix that may be
parametrized as

\be
{\cal M} =
\left(\begin{array}{cc}
-\mzchi\cosbchi\sinwchi &  \mzchi\sinbchi\sinwchi  \\
 \mzchi\cosbchi\coswchi & -C^{\chi}\mzchi\sinbchi\coswchi
\end{array}
\right) \ .
\ee
There are then seven parameters that enter neutralino events, and one
must try to check the SUSY relations $\mzchi = \mz$, $\theta_W^{\chi} =
\theta_W$ and $C^{\chi} = 1$.  A general analysis is likely to be
complicated.  One possible simplification would be to consider a less
than fully general neutralino mass matrix by setting, for example,
$C^{\chi} = 1$. On the other hand, one might wish to assume the
standard SUSY neutralino mass matrix, generalize the
neutralino-fermion-sfermion coupling to $\gchizero$, and check that
$\gchizero=g$.  However, even this analysis is more complicated than
the chargino case, because the SUSY neutralino mass matrix contains an
additional parameter.  In addition, an important caveat to all analyses
based on the neutralino mass matrix is that such analyses rely on the
absence of gauge singlets, and are therefore more model-dependent than
the chargino analyses of previous sections.

Without detailed study, it is not possible to dismiss the possibility
that precision studies of sfermion and neutralino properties may be
useful for testing SUSY. However, even from the brief comments
presented above, it is clear that the sfermion and neutralino sectors
are significantly less promising than the chargino sector. Category
(III) tests from chargino properties are likely to be the least model
dependent and may be the first strong quantitative tests even if some
other sparticles are lighter than charginos.

\section{Conclusions}
\label{sec:Conclusions}

Softly broken supersymmetric theories are like spontaneously broken
gauge theories in that the relationships between dimensionless
couplings implied by the symmetry continue to be preserved, while the
corresponding relationships between the masses of various particles can
be badly violated.  It is this feature which provides the best
opportunity for quantitative tests of supersymmetry.  In this study we
have examined the possibilities for testing various SUSY relations in a
number of scenarios. These studies have been conducted in the
experimental setting provided by a linear $\epem$ collider with
polarizable beams, and results have been presented for $\sqrt{s} = 500
\gev$ and integrated luminosities of 30 and 100 $\ifb$.

In the scenario in which charginos are the first sparticles to be
discovered, we have analyzed two representative cases. In the first, we
probed the form of the chargino mass matrix, and in the second, we
tested the $\chargino f \tilde{f}$ coupling.  In both examples, we
found that the test led to rather strong quantitative confirmations of
the MSSM and SUSY.  As a by-product, interesting bounds on some SUSY
parameters were also obtained. The availability of polarizable beams
was found to play a vital role, allowing us to define characteristic
regions, effectively eliminate dependences on certain SUSY parameters,
and remove background.  Our analysis was performed using a parton level
Monte Carlo event generator and did not incorporate possible
contamination of chargino pair events from other SUSY processes. Of
course, a more detailed analysis that includes the simultaneous
production of all possible SUSY events together with a more realistic
simulation is needed before definitive conclusions about precision SUSY
tests may be drawn.

The prospects for obtaining the first quantitative tests of SUSY from
sfermion and neutralino properties were also considered.  Sleptons were
found to be poor candidates for such tests because of the difficulty in
detecting sneutrinos, and precision tests from squarks were found to
rely on the discovery of at least four squarks or two squarks and, most
likely, two Higgs bosons. The analysis of neutralino properties is
complicated by its dependence on a large number of parameters.  Whether
these complications may be overcome in certain scenarios remains to be
seen in further studies. However, while falsification of sfermion mass
relations is the least model-dependent disproof of SUSY, it is likely
that the chargino sector is the simplest and most powerful for
verifying the quantitative predictions of SUSY.

We have not considered the possibilities for quantitative SUSY tests at
other colliders, nor have we examined the additional constraints that
come with the adoption of GUT and supergravity assumptions. Even with
fairly weak assumptions, however, we have found that, if sparticles are
produced at future $\epem$ colliders, measurements of their properties
may allow us to quantitatively verify SUSY, a valuable first step in
the exploration of the full structure of supersymmetric theories.

\acknowledgements

The authors thank M.~Drees, H.~Haber, T.~Rizzo, M.~Strassler, and
L.~Susskind for enlightening conversations. H.M. was supported by the
Director, Office of Energy Research, Office of High Energy and Nuclear
Physics, Division of High Energy Physics of the U.S. Department of
Energy under contract DE--AC03--76SF00098.

\appendix*{Uncertainty Analysis}

In this study we use the truncated forward-backward asymmetry $\achi
\equiv \achi_i$, where $i=L$ or $R$, and the $Y$ mode partial cross
section $\sigy$ to constrain parameter space. These theoretical
quantities are found through their correlations to experimental
observables. The uncertainties in determining $\achi$ and $\sigy$
therefore receive contributions from two sources: systematic errors,
that is, uncertainties arising from the lack of perfect correlation
between the theoretical quantities and the experimental observables,
and experimental statistical errors. In this appendix we collect the
formulae used to estimate the systematic and statistical errors.

Systematic errors are determined by performing Monte Carlo simulations
at a number of points in parameter space. The truncated
forward-backward asymmetry of chargino production before cuts, $\achi$,
is determined through its correlation to $\afbhad$, the
forward-backward asymmetry of the hadronic system's direction after
cuts.  The theoretical quantity $\achi$ depends only on parameters that
enter the production process, while $\afbhad$ depends on both
production and decay, and on cuts and detector effects.  The systematic
uncertainty in $\achi$ is therefore determined by the sensitivity of
$\afbhad$ to the decay process, cuts, and detector effects, and this
sensitivity is measured through simulations. For each of $\npts$ points
in parameter space, $\achi$ is determined from exact analytical
expressions, and $\afbhad$ is found from a Monte Carlo simulation. A
linear fit to the resulting distribution in the $(\achi,\afbhad)$ plane
is parametrized by

\be\label{fit}
\afbhad = a \achi + b \pm \amctot \ ,
\ee
where $\amctot$ is the $\onesigma$ uncertainty in $\afbhad$ for a fixed
$\achi$.  The total Monte Carlo uncertainty $\amctot$ includes both the
systematic error and fluctuations from finite Monte Carlo statistics.
The contribution from Monte Carlo statistical fluctuations is

\be\label{MC}
\amcstat = \left[ \frac{1}{\npts} \sum_{i=1}^{\npts} \left( \Delta
A_i^{\rm had} \right) ^2 \right]^{\frac{1}{2}} \ ,
\ee
where

\be
\Delta A_i^{\rm had} = \sqrt{\frac{1-\left(A_i^{\rm had}\right)
^2} {{N_{\rm MC}}_i}}
\ee
is the Monte Carlo statistical uncertainty in $A_i^{\rm had}$ for
simulation $i$, and ${N_{\rm MC}}_i$ is the effective number of events
in simulation $i$. The systematic error of the distribution is then

\be
\Delta A_{\rm exp}^{\rm sys} = \sqrt{
\left(\Delta A_{\rm MC}^{\rm tot}  \right)^2 -
\left(\Delta A_{\rm MC}^{\rm stat} \right)^2} \ .
\ee

To the systematic error must be added the experimental statistical
error.  This error is given by

\be\label{exp}
\Delta A_{\rm exp}^{\rm stat} = \sqrt{\frac{1-(\afbhad) ^2}{N_{\rm
exp}} + \frac{(1-\afbhad)^2}{N_{\rm exp}} \frac{N_{\rm back}}{N_{\rm
exp}}} \ ,
\ee
where $\afbhad$ is the forward-backward asymmetry for our case study,
and $N_{\rm exp}$ ($N_{\rm back}$) is the number of signal (background)
events that pass all cuts and is proportional to the integrated
luminosity. (Here we have assumed that the background is
well-understood and may be subtracted up to statistical uncertainties.
We also assume that all background events are in the forward
hemisphere, a good approximation for the dominant background, $W$ pair
production.) We estimate the total experimental uncertainty in
$\afbhad$ for a given $\achi$ to be

\be
\Delta A_{\rm exp}^{\rm tot} = \sqrt{
\left(\Delta A^{\rm sys}  \right)^2 +
\left(\Delta A_{\rm exp}^{\rm stat}\right)^2} \ .
\ee
What we actually measure is $\afbhad$, however.  We therefore are more
interested in the experimental uncertainty in $\achi$ for a fixed
$\afbhad$, which is

\be
\Delta A^{\chi} = |a|^{-1} \Delta A_{\rm exp}^{\rm tot} \ ,
\ee
where $a$ is the slope of the linear fit in Eq.~(\ref{fit}).

The efficiency of the cuts $\eta$ is found simply by its correlation to
previous measurements.  To determine the uncertainty in $\eta$, we
reduce the parameter space to the region in which the previous
measurements have their appropriate values and determine the variation
of $\eta$ within this subspace. We determine $\eta$ for each of the
simulations and obtain a distribution of points in the $(\achi,\eta)$
plane.  The best linear fit to this distribution is

\be
\eta = a' \achi + b' \pm \emctot \ ,
\ee
where $\emctot$ is the $\onesigma$ error in $\eta$ for a fixed $\achi$.
To find the systematic error, we must again remove the fluctuations
that arise solely from finite Monte Carlo statistics. The Monte Carlo
statistical error is

\be\label{demcstat}
\emcstat = \left[ \frac{1}{\npts} \sum_{i=1}^{\npts} \left( \Delta
\eta_i \right) ^2 \right]^{\frac{1}{2}} \ ,
\ee
where the statistical error for simulation $i$ is given by

\be
\Delta \eta_i = \sqrt{\frac{\eta_i (1-\eta_i )}{N_{\rm MC}}} \ .
\ee
The systematic error in $\eta$ for a fixed $\achi$ is then

\be\label{desys}
\esys = \sqrt{ \left(\emctot  \right)^2 - \left(\emcstat \right)^2} \ .
\ee
However, as seen above, $\achi$ is not determined exactly. The
uncertainty in $\achi$ weakens the determination of $\eta$, and the
total uncertainty in $\eta$ is

\be
\Delta \eta = \sqrt{\left(a' \Delta \achi \right)^2 + \left(\Delta
\eta^{\rm sys} \right)^2} \ .
\ee
We must now convert the uncertainty in $\eta$ into an uncertainty in
$\sigy$.  The $Y$ mode partial cross section and its fractional
uncertainty are given by

\be\label{sigmai}
\sigy = N_{\rm exp} \eta^{-1} {\cal L}^{-1}
\ee
and

\be\label{deltasigmay}
\frac{\Delta\sigy}{\sigy}
= \left[ \left( \frac{\Delta N_{\rm exp}}{N_{\rm exp}} \right)^2
+ \left( \frac{\Delta\eta}{\eta} \right)^2
+ \left( \frac{\Delta{\cal L}}{\cal L} \right)^2 \right]^{\half} \ ,
\ee
where $\cal L$ is the integrated luminosity.  For the purposes of this
study, $\Delta{\cal L}/{\cal L}$ is negligible. The uncertainty in the
number of $Y$ events passing the cuts is $\Delta N_{\rm exp} =
\sqrt{N_{\rm exp} + N_{\rm back}}$, where $N_{\rm exp}$ ($N_{\rm
back}$) is the number of $Y$ mode (background) events passing the cuts,
respectively, and we have again assumed that the background is
well-understood and may be subtracted up to statistical uncertainties.

\figure{\label{fig:productiondiagrams}
The diagrams contributing to chargino production at $e^+e^-$ colliders.
The $\snu_e$ $t$-channel diagram is absent for $e^-_R$ beams.}

\figure{\label{fig:sigmar}
Contours of constant $\sigma_R$ (in fb) for fixed $\tanb=4$ in the
$(\mu, M_2)$ plane.  Chargino production is inaccessible for $\sqrt{s}
= 500\gev$ in the hatched region, and the cross-hatched region is
excluded by the current experimental mass limit $\mchargino > 45 \gev$.
The cross section $\sigma_R$ quickly drops to zero in the $|\mu| \alt
M_2$ regions.}

\figure{\label{fig:regions}
The three characteristic regions for fixed $\tanb=4$ in the $(\mu,
M_2)$ plane, as defined in the text. (The corresponding $\mu>0$ parts
of these regions are unlabeled.) The hatched and cross-hatched regions
are as in Fig.~\ref{fig:sigmar}.  The dashed curve is the contour
$\mchargino = 172 \gev$.}

\figure{\label{fig:afb}
The correlation of $\afbhad$ and $\afbr$ for 38 points in the
seven-dimensional parameter space $(\mu, M_2, \tanbchi, M_1, \mslep,
\msq, \mwchi)$. These points have been picked randomly, subject only to
the constraints that $\mchargino$, $\mLSP$, and $\mcharginotwo$ are
within 2 GeV of their underlying values in the case study. The linear
best fit is given by the solid line, and the 1$\sigma$ deviations are
given by the dashed lines.}

\figure{\label{fig:eta}
The correlation of $\eta$ and $\afbr$ for the 38 points in the
seven-dimensional parameter space $(\mu, M_2, \tanbchi, M_1, \mslep,
\msq, \mwchi)$, selected as in Fig.~\ref{fig:afb}.  The linear best fit
is given by the solid line, and the 1$\sigma$ deviations are given by
the dashed lines.}

\figure{\label{fig:phiplane}
The allowed region of the $(\phi_+, \phi_-)$ plane from measurements of
$\afbr$ and $\sigy$.  The lightly (heavily) shaded region is allowed
for ${\cal L}_R = 30 \, (100) \, \ifb$. Contours of constant $\mwchi$
are plotted in GeV. On the dotted contours, the SUSY relation $\mwchi =
\mw$ holds.}

\figure{\label{fig:msnugchiplane}
Allowed regions (shaded) of the $(\msnu, \gchi)$ plane for ${\cal L}_L$
= (a) 30 $\ifb$ and (b) 100 $\ifb$. Solid (dashed) curves are contours
of constant $\sigma_L$ ($\afbl$) that bound the allowed regions.  On
the dotted lines, the SUSY relation $\gchi = g$ is satisfied.}

\end{document}